\documentclass[letterpaper, 10 pt, conference]{ieeeconf}

\IEEEoverridecommandlockouts                              

\usepackage[pdftex]{graphicx}
\usepackage{siunitx}
\usepackage{subfig}
\usepackage{cancel}
\usepackage{mathtools}
\usepackage{amsmath}
\usepackage{amssymb}
\usepackage{siunitx}
\title{\LARGE \bf
	Time regularization as a solution to mitigate quantization induced performance degradation
}

\author{Bas Kieft, S. Hassan HosseinNia$^{*}$ and Niranjan Saikumar%
	\thanks{All authors are with Precision and Microsystems Engineering, 3ME, Delft University of Technology, The Netherlands}%
	\thanks{$^{*}$Corresponding author - \small{S.H.HosseinNiaKani@tudelft.nl}}%
	\thanks{The authors would like to thank Hittech Multin. This work was supported by NWO, through OTP TTW project $\#16335$.}
}

\begin{document}
	
	\maketitle

\begin{abstract}
Reset control is known to be able to outperform PID and the like linear controllers. However, in motion control systems, quantization can cause severe performance degradation. This paper shows the application of time regularization to mitigate this practical issue in reset control systems. Numerical simulations have been conducted in order to analyze the cause of the quantization induced performance degradation and the effectiveness of time regularization to mitigate this degradation; with tuning guidelines for the time regularization parameter also provided. Moreover, a robustness analysis is performed. The solution is also tested experimentally on a high precision motion system for validation. It is estimated by numerical simulations that time regularization can reduce quantization induced performance degradation by almost \SI{10}{dB}. Experiments have similarly shown a reduction of several dB for the high precision motion stage.
\end{abstract}


\section{Introduction}
Over the past years, reset control has found increasing popularity as a complement to linear Proportional-Integral-Differential (PID) control. This is because the reset action allows the controller to surpass the linear limitations as posed by Bode. Reset control was introduced by Clegg in the form of a Clegg Integrator (CI) \cite{Clegg2013}. CI is an integrator whose state is reset to zero when the input is zero. Thus, the phase lag is reduced from $\SI{-90}{^\circ}$ to $\SI{-38}{^\circ}$. The First Order Reset Element (FORE) was created as a continuation in reset control as a nonlinear low-pass filter (LPF) \cite{Horowitz1975}, while the Second Order Reset Element (SORE) is a second order nonlinear LPF \cite{Hazeleger2016}.

Methods to tune the reset nonlinearity have been developed in the form of PI+CI control and partial reset. A CI parallel with a PI results in a PI+CI element, and with weighted gains in each branch allows for additional tuning freedom \cite{Ba2007}. Similarly, while in a full reset controller, the state is reset to 0, partial reset allows the state to be reset to a predetermined fraction of its value prior to reset. This when done with a FORE creates a Generalized FORE (GFORE) \cite{Guo2009} and with SORE creates a Generalized SORE (GSORE) \cite{Saikumar2019}. Fractional order elements in reset control was advanced in \cite{Saikumar2017} with the introduction of the Generalized Fractional Order Reset Element (GFrORE).

Reset control has found numerous applications in both the process industry and motion control. A PI+CI element was applied to an industrial heat exchanger in \cite{Vidal2010}. Furthermore, this paper researches the application of a variety of reset conditions. In \cite{Li2011a}, \cite{Li2009a} and \cite{Guo2009}, Hard Disk Drives (HDD) are used as a universal base for reset control research. Moreover, applications of reset control span the field from tape-speed setups \cite{Zheng2000} to exhaust gas re-circulation \cite{Panni2014}.

The resetting action results in reduced phase lag compared to their linear counterparts and is advantageous. However, more interesting elements can be created by cleverly combining reset and linear elements. The Constant in gain Lead in phase (CgLp) \cite{Saikumar2019, Palanikumar2018} element which combines a GFORE with a linear first-order lead element to achieve an increase in phase while not altering the gain is such an element. A second order CgLp is possible by implementing a GSORE element with a second-order linear lead. Since, the frequency domain approach of loop-shaping is the most widely used method in industry, Sinusoidal Input Describing Function (SIDF) is a common linearization used with reset controllers. This approach was adopted for all discussed elements \cite{Clegg2013,Horowitz1975,Hazeleger2016,Hosseinnia2013,Ba2007,Guo2009,Saikumar2019,Palanikumar2018}. 

In most cases in literature, results are extracted from simulations where continuous time implementations are used. In \cite{Palanikumar2018, Chen2019a, Saikumar2019}, the controllers were discretized and no significant performance deterioration is reported. However, quantization is a type of practical issue that is not studied in literature with the application of reset control.

Quantization is a form of signal distortion that is often compared to, or included in noise, or modeled as white or colored noise \cite{Zhu2013}. It has been researched extensively, ex. \cite{Kaufmann2012, Linnenbrink2006, Lai2003} and its influence on linear control is well known \cite{Ferrante2015, Minyue2008, Park2019}. While reset control has been shown to have superior noise attenuation properties due to the reduced high frequency gain \cite{Palanikumar2018, Chen2019, Saikumar2019}, the effects that quantization can have on reset control is not well studied. However, evaluation of reset controllers based on noise attenuation properties does not represent reality. For example, the frequency of distortion is dependent on the quantizer resolution and response slope. Some research in quantization combined with reset control was done in \cite{Zheng2010} where the state estimator had reset elements in order to mitigate quantization generated signal distortion. However, no works exist which study the effects and propose solutions to overcome any performance degradation.  

Due to the sharp nature of the resetting action and the nature of the resetting surface with the zero error crossing condition, time regularization was introduced as a method to overcome Zenoness \cite{Johansson1999a}. Zenoness is when a theoretical system gets stuck in time and cannot advance. In \cite{Nesic2008}, time regularization is used in order to avoid Zenoness in reset control. This approach uses a holding function on the reset condition such that no resets are induced within this holding period. In practical setups, Zenoness will not occur due to discretization, among other reasons. However, the nature of the resetting surface in the presence of quantization can similarly result in unmodelled resets. Hence, in this paper we study the use of time regularization to overcome the performance deterioration that occurs due to quantization.

The main contributions of this paper are a proposed solution for quantization induced problems based on the application of time regularization and a method to tune the proposed solution. The paper is structured as follows: Section \ref{sec:preliminaries} states the preliminaries required for the paper. Then in section \ref{sec:QIPD}, the performance degradation is expanded upon. Section \ref{sec:TimeReg} presents time regularization to overcome these issues with tuning rules also proposed. The results from the experimental setup are presented in section \ref{sec:Application} and the conclusions are provided in section \ref{sec:Conclusion}.

\section{Preliminaries}\label{sec:preliminaries}

\subsection{Reset Control Definition}
The most commonly used form of reset controller with the zero-error crossing condition can be represented as in (\ref{eq:resetdef}) \cite{Saikumar2019,Guo2009}.
\begin{equation}\label{eq:resetdef}
R: \; \; 
\begin{cases} 
\dot{\mathbf{x}}_r(t) = A_r\mathbf{x}_r(t) + B_r e(t) & \text{if } e(t) \ne 0\\
\mathbf{x}_r(t^{+})=A_\rho\mathbf{x}_r(t) &  \text{if } e(t) = 0\\
u(t)=C_r\mathbf{x}_r(t)+D_r e(t)
\end{cases}
\end{equation}
where $\mathbf{x}_r$ are the reset states, $u$ is the output, e is the error input, and $A_r$, $B_r$, $C_r$ and $D_r$ are the state-space matrices and are referred to as the base linear system. $A_\rho$ is diagonal and is the reset matrix that determines the states to be reset and their after reset value. For a full reset $A_\rho$ is the zero matrix and for a linear controller, this is the identity matrix.

\subsection{Describing Function}\label{sec:DF}
Because reset control is nonlinear, a Sinusoidal Input Describing Function (SIDF) analysis can be utilized for approximation in the frequency domain. In \cite{Guo2009} the describing function for the reset element of (\ref{eq:resetdef}) is provided as:
\begin{equation}\label{eq:G}
G(j\omega)=C^{T}_r(j\omega I-A_r)^{-1}(I+j\Theta_\rho(\omega))B_r+D_r
\end{equation}
where 
\begin{equation}\label{eq:Theta}
\Theta_\rho \overset{\Delta}{=} \frac{2}{\pi} \left( I+e^{\frac{\pi A_r}{\omega}}\right)\left(\frac{I-A_\rho}{I+A_\rho e^{\frac{\pi A_r}{\omega}}}\right)\left(\left(\frac{A_r}{\omega}\right)^{2}+I\right)^{-1}
\end{equation}

\subsection{Stability}
The stability of reset control has been investigated and necessary conditions termed as the $H_\beta$ condition can be found in \cite{Beker2002}. A Reset control system (RCS) for a plant defined with matrices $A_p, B_p, C_p \text{ and } D_p$, and with a reset controller defined in (\ref{eq:resetdef}) as its feedback controller is quadratically stable, if there exists a $\beta \in \mathbb{R}^{n_r \times 1}$ and a positive definite $P_\rho \in \mathbb{R}^{n_r \times n_r}$ such that
\begin{align}
	\mathcal{H}_\beta \triangleq 
	\begin{bmatrix}
		\beta C_P & 0_{n_r \times n_{nr}}& P_\rho
	\end{bmatrix} \left(sI - A_{cl}\right)^{-1}
	\begin{bmatrix}
		0 \\ 0_{n_{nr} \times n_r}\\I_{n_r}
	\end{bmatrix}
\end{align}
is strictly positive real, with
$A_{cl}=\begin{bmatrix}
	A_p & B_p C_r \\
	-B_r C_p & A_r
\end{bmatrix}$\\
as the closed-loop $A$-matrix. $n_r$ indicates the number of states being reset, $n_{nr}$ indicates the number of non-resetting states. For partial reset, an additional condition given below needs to be satisfied.
$$A_\rho P_\rho A_\rho - P_\rho \leq 0$$

\subsection{Reset Elements and select controller designs}
\subsubsection{Clegg integrator}
The CI was presented in 1958 \cite{Clegg2013}. Following the definition of (\ref{eq:resetdef}), the CI is defined with $A_r = 0$, $B_r = 1$, $C_r = 1$, $D_r = 0$ and $A_\rho = 0$.  

\subsubsection{FORE}
To offer a tunable frequency range of nonlinearity, the First Order Reset Element (FORE) was presented as a non-linear low-pass filter (LPF) with corner frequency $\omega_r$ \cite{Horowitz1975}. Following the definition of (\ref{eq:resetdef}), the FORE is defined with $A_r = -\omega_r$, $B_r = \omega_r$, $C_r = 1$, $D_r = 0$ and $A_\rho = 0$. A generalisation for FORE (GFORE) has been defined in \cite{Guo2009} where $A_\rho = \gamma$ with $-1\le\gamma\le1$.  $\gamma =1$ represents a base linear first order low-pass filter.

\subsubsection{SORE}
The Second Order Reset Element (SORE) was introduced in \cite{Hazeleger2016} with the following matrices:
\[A_r=\begin{bmatrix}
0 & 1 \\
-\omega_r^2 & -2\beta_r \omega_r
\end{bmatrix}, 
B_r = \begin{bmatrix}
0\\
\omega_r^2
\end{bmatrix},\]
\[C_r=\begin{bmatrix}
1 & 0
\end{bmatrix},
D_r=\begin{bmatrix}
0
\end{bmatrix}.\]
where $\beta_r$ is the damping coefficient. SORE has been adjusted to allow for partial reset in the GSORE element \cite{Saikumar2019} resulting in $A_\rho = \gamma I$, where $-1\le\gamma\le1$.

\subsubsection{Constant in gain Lead in phase element}
CgLp is designed using a FORE/SORE element in combination with a corresponding first/second order linear lead. The state space representation of first order CgLp is with matrices:
\[
A_r=  \begin{bmatrix}
-\omega_{r\alpha} & 0\\
\omega_f & -\omega_f 
\end{bmatrix}, \quad
B_r=\begin{bmatrix}
\omega_{r\alpha}\\
0
\end{bmatrix},\]
\[
C_r=\begin{bmatrix}
\dfrac{\omega_f}{\omega_r} & \left(1-\dfrac{\omega_f}{\omega_r}\right)
\end{bmatrix},\quad
D_r=\begin{bmatrix}
0
\end{bmatrix},\quad
A_\rho = \begin{bmatrix}
\gamma & 0\\
0 & 1
\end{bmatrix}.
\]
where $\omega_{r\alpha}$ , $\omega_f$ and $\omega_r$ are the corner frequency of FORE, starting and taming frequencies of the linear lead respectively. With CgLp, $\omega_{r\alpha}$ and $\omega_r$ are close to each other, resulting in a gain cancellation, but with phase lead in the [$\omega_r$, $\omega_f$] range.

\subsection{CgLp-PID}
The CgLp element is used in combination with a PID controller in series. The design procedure is stated in \cite{Saikumar2019}. As per the $H_\beta$ stability condition provided earlier, the base linear system needs to be stable. CgLp, having the constant gain but lead in phase, can be used to partially or completely replace the D part of PID. A CgLp-PID can be represented as:
\begin{equation}\label{eq:CgLp-PID}
C=\underset{PI}{\underbrace{K(1+\frac{\omega_i}{s}})}\underset{D}{\underbrace{\frac{(s/\omega_d+1)}{(s/\omega_t+1)} }}\underset{CgLp}{\underbrace{\frac{1}{\cancelto{\gamma}{{s}/{\omega_{r\alpha}}+1}}\;\;\;\;\frac{{s}/{\omega_r +1}}{{s}/{\omega_f+1}} }}
\end{equation}
where the arrow indicates resetting action on the associated filter. The phase lead provided by D can be tuned by varying $a$ in $\omega_d = \omega_c/a$ and  $\omega_t = \omega_c a$. The phase lead provided by CgLp can be tuned by $\omega_r$ and $\gamma$. 

\section{Quantization induced performance deterioration in reset control systems}\label{sec:QIPD}
\begin{figure}[h!]
	\centering
	\includegraphics[width=1\columnwidth]{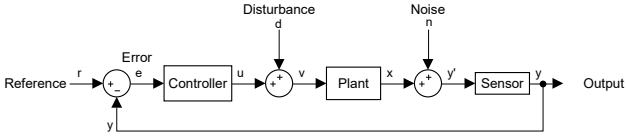}
	\caption{Block diagram of a feedback controller system showing the addition of distortion through quantization by the sensing system.}\label{fig:BD}
\end{figure}

Consider the closed loop control structure as shown in Fig. \ref{fig:BD} including the noise picked up the sensor and the additional distortion caused by quantization.

\subsection{Distortion}
The error associated with quantization is known as distortion \cite{Linnenbrink2006}. Due to its abundance in industrial applications, rounding quantization will be followed in this paper. A quantization level is defined as $Q=\frac{Range}{2^{bits}}$, where range is the maximum value that can be detected and bits is related to the digitization of the data and is the number of bits available to show the magnitude of the value. As an example, for a sensor with a range of $\SI{1000}{nm}$, 5 bit and 6 bit systems will result in a resolution and $Q$ of $\frac{\SI{1000}{nm}}{2^5}=\SI{31.25}{nm}$ and $\frac{\SI{1000}{nm}}{2^6}= \SI{15.625}{nm}$ respectively. The boundary between two levels of quantization in the sensor signal can be referred to as a jump or transition level. 

\subsection{Performance deterioration} \label{sec:RQLC}
To show that distortion causes performance degradation in reset systems, a mass based positioning system with $m=\SI{1}{kg}$ and controller with parameter values provided in Table \ref{tab:Settingsmass} are used. A mass system has no resonances and has predictable trend for the sensitivity function ($S$). Hence, any aberration indicates deviated controller behavior allowing for easier analysis. A linear sensitivity function does not fully hold due to the presence of the nonlinear reset controller in the loop. Hence $S_\sigma$ as proposed by \cite{Cai} and defined in (\ref{eq:Ssigma}) is used.
\begin{align}
\label{eq:Ssigma}
&S_{\sigma}(f)=\frac{\max (|e(t)|)}{|r|}
&\text { for } t \geq t_{s s}
\end{align}
where $r=A\sin{(2\pi ft)}$.

\begin{table}[tbp]
	\centering
	\caption{Controller settings applied to the mass stage.}
	\label{tab:Settingsmass}
	{
		\begin{tabular}{|c|c|c|}
			\hline
			K                  & \SI{6.0954e+05}{}  & -     \\ \hline
			$\omega_c$         & \SI{942}{}      & \SI{}{rad/s}    \\ \hline
			$\omega_i$         & \SI{94}{}       & \SI{}{rad/s} \\ \hline
			$\omega_d$         & \SI{530}{}      & \SI{}{rad/s} \\ \hline
			$\omega_t$         & \SI{1.68e+03}{} & \SI{}{rad/s} \\ \hline
			$\omega_{r\alpha}$ & \SI{160}{}      & \SI{}{rad/s} \\ \hline
			$\omega_f$         & \SI{9.42e+03}{} & \SI{}{rad/s} \\ \hline
			$\omega_r$         & \SI{172}{}      & \SI{}{rad/s} \\ \hline
			$\gamma$           & \SI{0.5}{}      & -     \\ \hline
			Range              & \SI{5000}{}     & $\SI{}{\mu m}$    \\ \hline
			$F_s$			   & \SI{10}{}       & \SI{}{kHz}    \\ \hline
		\end{tabular}%
	}
\end{table}

\begin{figure}
	\centering
	\includegraphics[width=1\columnwidth]{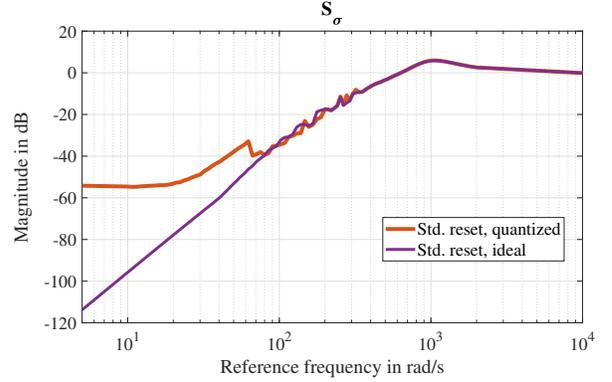}
	\caption{$S_\sigma$ for a mass system with and without quantiation. In the case of quantization, the maximum resolution is $\SI{9.8}{\mu m}$.}\label{fig:Ssigmar0}
\end{figure}

\begin{figure}
	\centering
	\includegraphics[width=1\columnwidth]{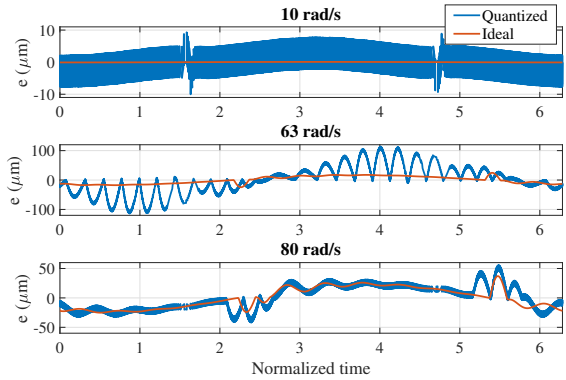}
	\caption{One period of the steady state error response for systems with and without quantization.}\label{fig:lastper}
\end{figure}

Fig. \ref{fig:Ssigmar0} shows the $S_\sigma$ for the considered mass based positioning system. While ideally the linear sensitivity function $S$ as well as the here considered $S_\sigma$ go to zero as frequency tends to zero, when quantization is at play, the steady state error can never go below $Q$. This leads to an altered $S$ that has a theoretical limit at $|\frac{E}{Q}|$, where $E=1$ for a normalized error input. Fig. \ref{fig:lastper} shows one period of the steady state error for a reference input of frequency \SI{10}{rad/s}.This effect of the minimum limit on error due to quantization can be seen in this case. However, this is a limitation also seen in linear systems.

However, the response for \SI{63}{rad/s} shows a significant difference in response between the quantized and ideal systems, and this is specific to reset control. This is due to quantization induced excessive resetting. In Fig. \ref{fig:Ssigmar0}, this shows as a break in the expected +2 slope trendline and a bump can be seen. Again at higher frequencies, where the dynamics are faster, quantization does not introduce unnecessary resets and the difference between ideal and quantized feedback signal is minimal as seen in Fig. \ref{fig:lastper} for \SI{80}{rad/s}.

When reviewing (\ref{eq:resetdef}), it can be deduced that due to the nature of the resetting surface, a quantization jump which has a magnitude of $\frac{1}{2}Q$ with respect to the mean of the error, can induce reset when the mean of the error is within $\frac{1}{2} Q \ge e \ge -\frac{1}{2} Q$. Ideally, the reset condition is satisfied when the error is zero. However, since quantization leads to resetting in a small band, multiple unmodelled resets occur resulting in significant increase in error as seen.

\section{Time regularization as solution for quantization induced performance degradation}\label{sec:TimeReg}
\subsection{Time regularization preliminaries}
Performance deterioration seen with quantization occurs with excessive resetting. In literature a similar phenomenon is discussed, known as Zenoness \cite{Banos2012}. A Zeno-solution is when an infinite number of discrete transitions take place in a finite amount of time. Although Zeno solution does not occur in practice due to discretization of the controller, the solution proposed in \cite{Johansson1999a} is useful to solve the problem seen with quantization. Quantization causes performance deterioration by initiating an excessive amount of resets in a tight sequence. Hence, a holding time will prevent subsequent resets and can be expected to improve performance. 

This paper brings novelty to the time regularization field by showing that it can mitigate quantization induced performance degradation. Moreover, while time regularization can be achieved with any finite time interval to avoid Zeno, this is not true for its use to mitigate quantization induced deterioration. Hence, we provide tuning rules for the same as well.

\subsubsection{Definition}
A generally adopted definition of time regularization can be found in (\ref{eq:TR2}) from \cite{Banos2012}.
\begin{equation}\label{eq:TR2}
R_{\rho}: 
\begin{cases} 
\dot{\mathbf{x}}_r(t) = A_r\mathbf{x}_r(t) + B_r e(t), & \dot{\tau}=1  \\&\text{if } e(t) \ne 0 \cup \tau \leq \rho \\
\mathbf{x}_r(t^{+})=A_\rho\mathbf{x}_r(t), &  \tau^{+}=0 \\ &  \text{if } e(t) = 0 \cap \tau > \rho\\
u(t)=C_r\mathbf{x}_r(t)+D_r e(t)
\end{cases}
\end{equation}
where $\tau$ is a counter that starts counting after a reset and $\rho$ is a tunable value that lower limits the time between resets.

\subsubsection{Sinusoidal Input Describing Function}\label{sec:TRDF}
The SIDF describes the response of an element based purely on a sinusoidal input. For the SIDF as obtained in the absence of time regularization to be valid with the same, there can be no missed resets. A standard reset system should reset twice per period, so the holding time cannot be longer than half a period of the input sine. Therefore the DF as provided in (\ref{eq:G}) is only valid up to:
\begin{equation}\label{eq:dfmax}
\omega_{DF max} = \frac{\pi}{\rho}
\end{equation}

\subsubsection{Stability}
The stability theorem stated earlier is also applicable in this case as the proof depends on the reduction of the Laypunov function in the regular flow condition as well as the jump condition. Since, time regularization removes resets and does not by itself introduce additional resets, the system is still quadratically stable.	

\subsection{Tuning rules}\label{sec:ROT}
Consider a sinusoidal reference $r=A\sin{(2\pi ft)}$. Based on linear control the error will be a sinusoid $e=E\sin{(2\pi ft + \phi)}$ where $\phi$ is the phase shift.
The error will cross zero twice per period, resulting in two resets per period. The maximum frequency of the reference that a feedback controller has to be able to track is the crossover frequency. Therefore the largest holding time need not be bigger than half a period of the bandwidth frequency (defined as crossover frequency)
\begin{equation}\label{eq:rho}
\rho \le \frac{1}{2f_c}
\end{equation}

This will eliminate excessive resetting up to bandwidth and ensure the reliability of the DF in analysis and design. However, any changes in the system, for ex. due to gain variation might alter the cross-over frequency resulting in the holding time $\rho$ being too short for the new crossover frequency. Therefore a safety factor $k$ has been included to get
\begin{equation}\label{eq:RoT}
	\rho = \frac{1}{2k\cdot f_c}
\end{equation}
For shorthand the TR condition will be denoted as TR - $k\cdot f_c$.

\subsection{Numerical sensitivity function}
\begin{figure}
	\centering
	\includegraphics[width=1\columnwidth]{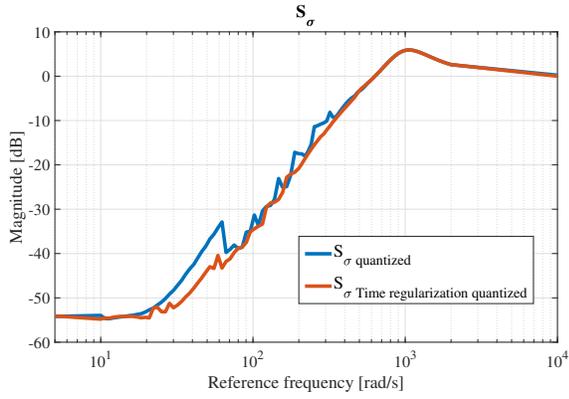}
	\caption{$S_\sigma$ with quantization for standard reset and TR - $\frac{5}{2} \cdot f_c$. }\label{fig:Sq}
\end{figure}

The same mass system as for earlier was considered to analyze the performance of the solution. Fig. \ref{fig:Sq} shows the performance improvement achieved by implementing time regularization according to the tuning guideline with $k=\frac{5}{2}$. In the region where quantization has the most influence for the standard reset condition and degrades tracking performance, an improvement of up to 10 dB is achieved.

\subsection{Sensitivity to $\rho$}\label{sec:tradeoff}
In order to evaluate the tuning sensitivity with respect to $\rho$, $S_\sigma$ was analyzed in simulation. According to (\ref{eq:dfmax}) and (\ref{eq:rho}), the holding time limits the describing function and applicability of DF for tuning up to a predetermined frequency. The peak of the sensitivity function is an important factor in control design. This peak occurs at frequencies slightly higher than the crossover frequency. This creates a trade off between lower frequency performance improvement and the high frequency sensitivity peak. A long holding time will prevent more excessive resets while also increasing the peak of the sensitivity function due to missed resets. Therefore, an analysis in robustness and performance has been performed by using a range of safety factors $k$.

\begin{figure}
	\centering
	\includegraphics[width=1\columnwidth]{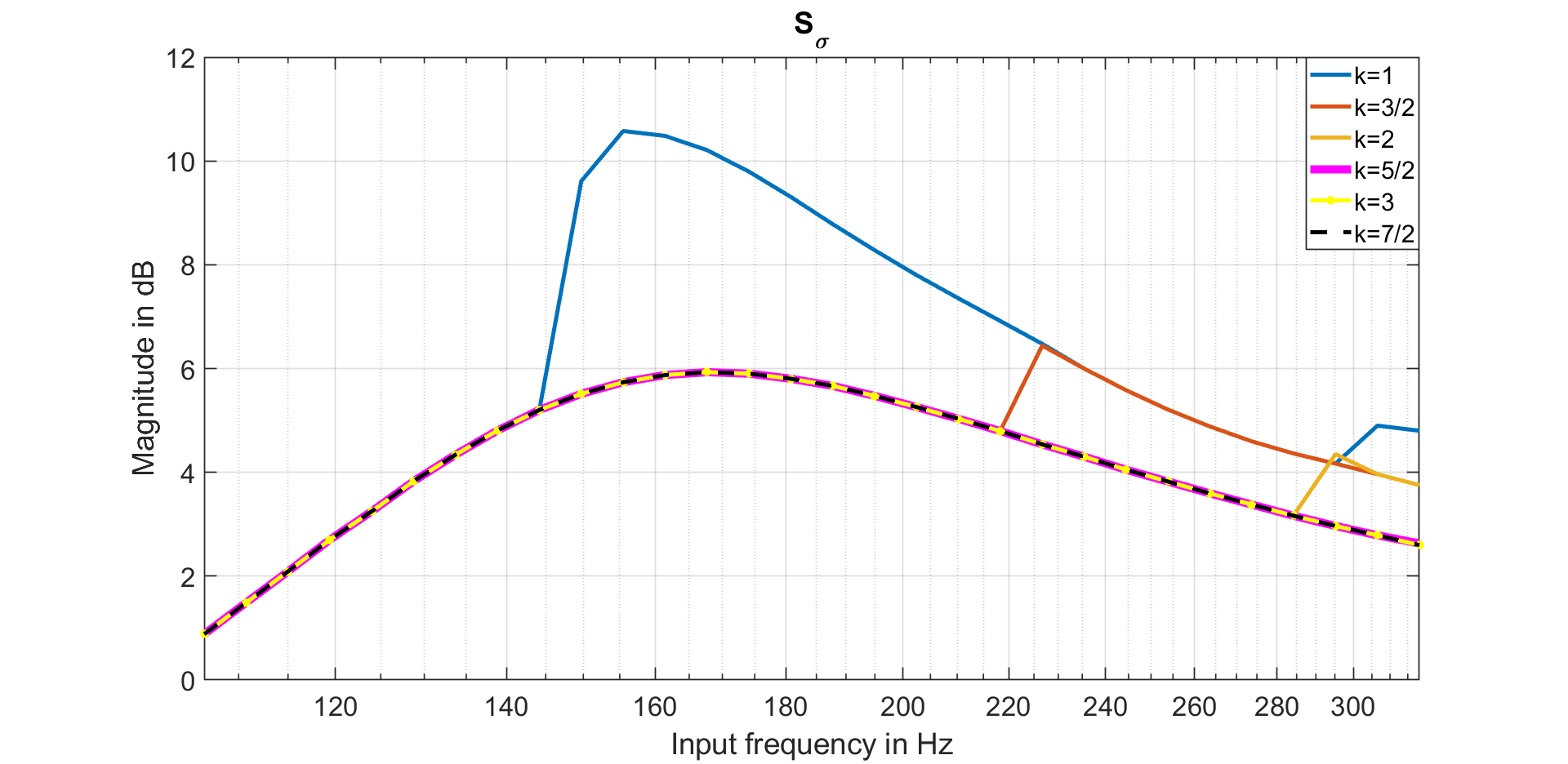}
	\caption{The peak of $S_\sigma$ for multiple values of k. }\label{fig:S_MM}
\end{figure}

The peak of $S_\sigma$ and the frequency range where improvement is desired are analyzed. Fig. \ref{fig:S_MM} shows the peak of $S_\sigma$ for multiple values of $k$. It can be seen that $k$ has a clear influence and $k$ should be sufficiently high such that this peak is not increased. Especially the jump at \SI{150}{Hz} for $\rho=\frac{1}{2kf_c}$ with $k=1$ clearly shows the effect of resets being missed due to time regularization. While this is useful to avoid unmodelled resets, missed resets due to the same solution can create an additional problem if not tuned correctly.

\begin{figure}
	\centering
	\includegraphics[width=1\columnwidth]{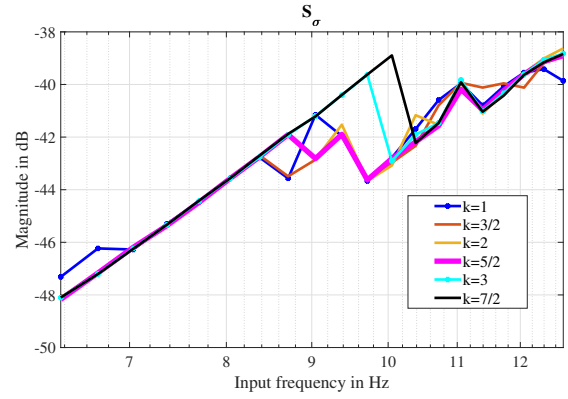}
	\caption{The frequency range of desired improvement of $S_\sigma$ for multiple values of k. }\label{fig:S_lf}
\end{figure}

In Fig. \ref{fig:S_lf}, the influence of $k$ on the frequency range of desired improvement is shown. The performance improves for increasing values of k, up to a limit. From Fig. \ref{fig:S_MM} it can be seen that $k=5/2$ results in the shortest holding time without increasing the peak of sensitivity function. Beyond which, the performance improvement is reduced again. Based on these observations of the simulation results, the following remark is made: \\
\noindent Concluding from the sensitivity and performance analysis $k=\frac{5}{2}$ is the advised safety factor.

\section{Application}\label{sec:Application}
\subsection{Precision positioning stage}
\begin{figure}
	\centering
	\includegraphics[width=1\columnwidth]{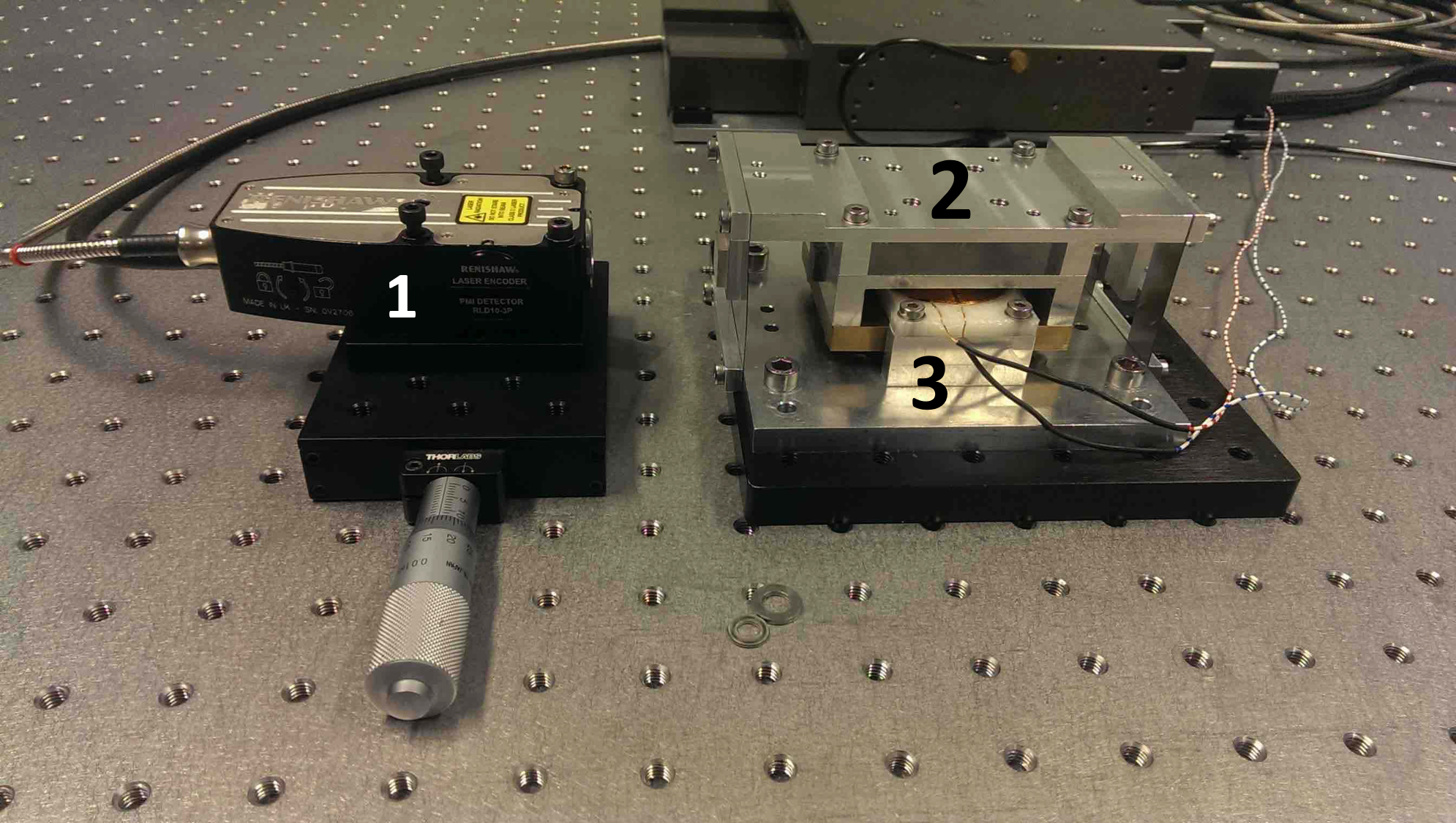}
	\caption{Fine positioning stage. 1: sensor, 2: mass stage, 3: actuator.}\label{fig:finestage}
\end{figure}

In order to validate the theory a custom designed one-degree-of-freedom high precision positioning stage has been used. In essence, it is a mass-spring-damper system. The stage can be seen in Fig. \ref{fig:finestage}. The sensor is a Renishaw RLE10 laser encoder set to \SI{10}{nm} resolution and the actuator is Lorentz force based. In order to achieve fast real-time control an FPGA NI cRIO system was utilized with a sampling rate of \SI{10}{kHz}.

\begin{figure}
	\centering
	\includegraphics[width=1\linewidth]{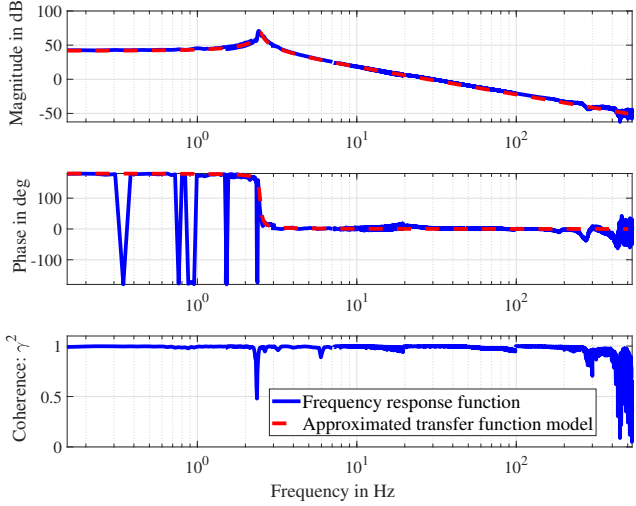}
	\caption{Frequency response function of the positioning system and Bode plot of the estimated model.}\label{fig:Tfestimate}
\end{figure}

Frequency response data of the system is obtained by applying chirp signals, as is common in industry. The results can be seen in Fig. \ref{fig:Tfestimate}. The following transfer function was estimated:
\[P(s)=\frac{3.038e04}{s^2 + 0.7413 s + 243.3}.\]

\subsection{Designed controllers}
It was decided to study CgLp-PID controller for a bandwidth, defined as crossover frequency, of \SI{150}{Hz} and phase margin of $\SI{40}{^\circ}$. The CgLp-PID was designed using the estimated transfer function of (\ref{eq:CgLp-PID}) with the controller parameters shown in Table \ref{tab:SettingsCgLp}. The controller was discretized according to the tustin method.
\begin{table}[]
	\centering
	\caption{Controller settings for practical application.}
	\label{tab:SettingsCgLp}
	\begin{tabular}{|c|c|c|}
		\hline
		K                  & \SI{16.41}{} & -     \\ \hline
		$\gamma$           & \SI{0}{}     & -     \\ \hline
		$\omega_c$         & \SI{942.5}{} & \SI{}{rad/s} \\ \hline
		$\omega_i$         & \SI{94.25}{} & \SI{}{rad/s} \\ \hline
		$\omega_d$         & \SI{529.2}{} & \SI{}{rad/s} \\ \hline
		$\omega_t$         & \SI{1679}{}  & \SI{}{rad/s} \\ \hline
		$\omega_{r\alpha}$ & \SI{697.6}{} & \SI{}{rad/s} \\ \hline
		$\omega_r$         & \SI{812.1}{} & \SI{}{rad/s} \\ \hline
		$\omega_f$         & \SI{9420}{} & \SI{}{rad/s} \\ \hline
	\end{tabular}
\end{table}

\subsection{Results}
In order to show that quantization induced performance degradation is not restricted to one specific $Q$, two $Q$'s have been studied. The highest resolution corresponds to \SI{10}{nm} and another where the resolution was artificially reduced to \SI{80}{nm}.

\subsubsection{\SI{10}{nm} resolution}
\begin{figure}
	\centering
	\includegraphics[width=1\columnwidth]{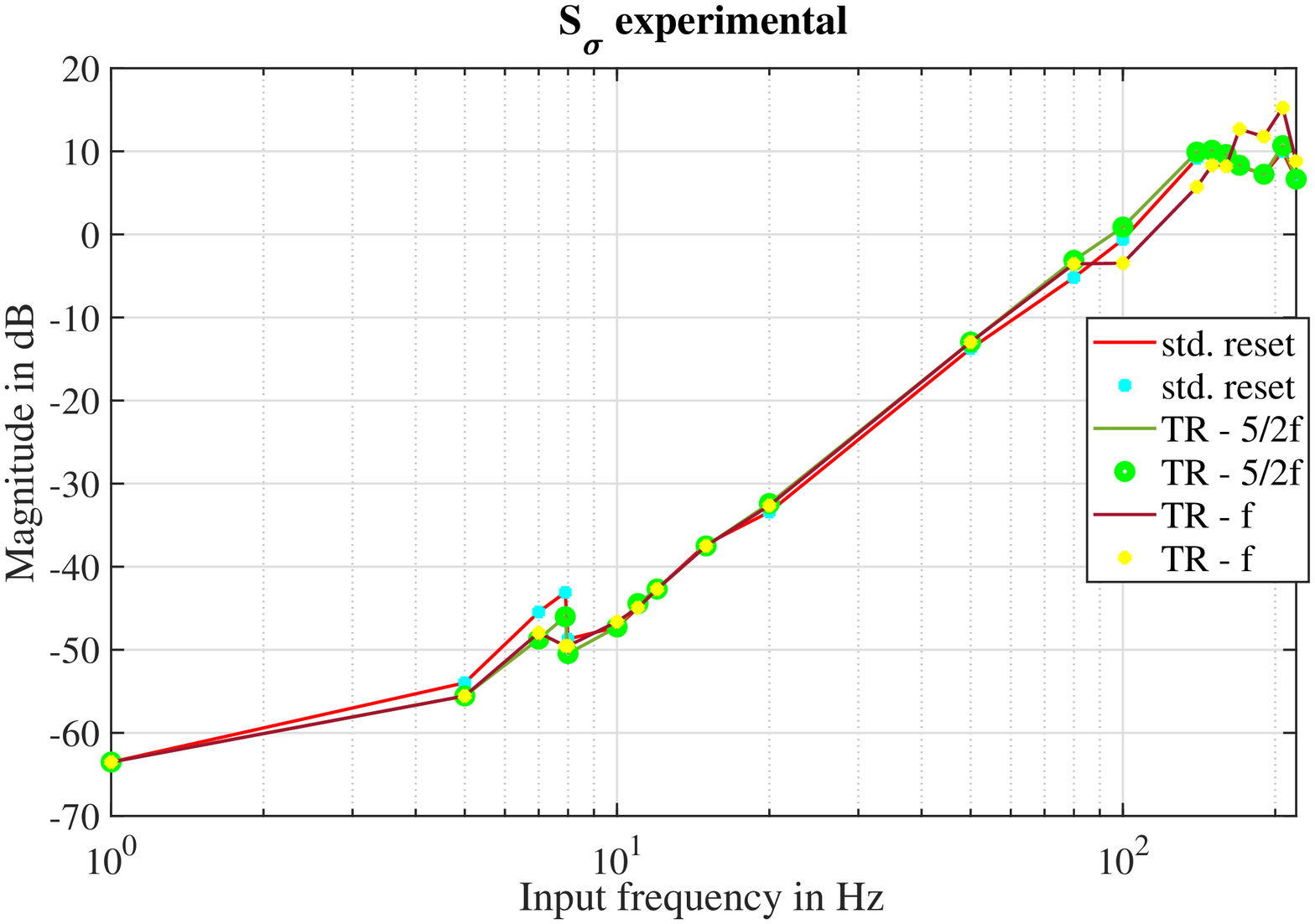}
	\caption{Experimentally deduced sensitivity function for CgLp-PID with and without time regularization (TR - $\frac{5}{2}\cdot f_c$ and TR - $f_c$). The sensor has a resolution of \SI{10}{nm}.}\label{fig:S_CgLp_exp_Q0}
\end{figure}

A reference amplitude of $\SI{30}{\mu m}$ was applied at low frequencies. The minimum possible error is \SI{10}{nm}, which leads to a limit of \SI{-69}{dB}. In practice this limit was not achieved due to noise and disturbance. Due to the limited linear stroke of the stage and saturation levels of the system, the amplitude of the applied reference was decreased for higher frequencies. For this quantization level, in combination with the controller and plant dynamics, the quantization induced performance degradation occurred in the lower frequency range. In Fig. \ref{fig:S_CgLp_exp_Q0}, a bump similar to Fig. \ref{fig:Ssigmar0} can be seen starting from \SI{5}{Hz} and ending at \SI{7.9}{Hz}. Time regularization based on $\rho = \frac{1}{2f_{c}}$ shows a significant improvement for quantization induced performance degradation. As expected, an increase in the sensitivity peak can be seen. Time regularization with $k=\frac{5}{2}$ shows an improvement in the low frequency domain while maintaining the peak in sensitivity function at the required levels.

\subsubsection{\SI{80}{nm} resolution}
\begin{figure}
	\centering
	\includegraphics[width=1\columnwidth]{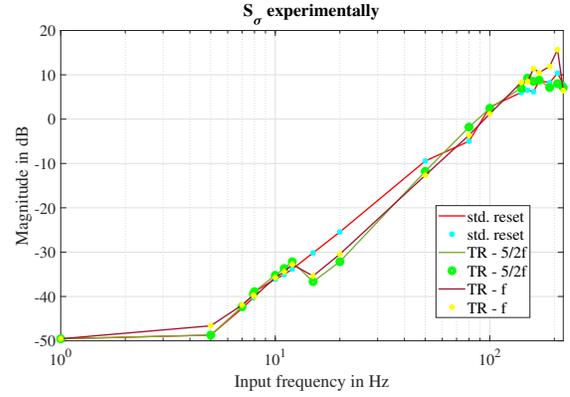}
	\caption{Experimentally deduced sensitivity function for CgLp-PID with and without Time Regularization (TR - $\frac{5}{2}\cdot f_c$ and TR - $f_c$). For 80 nm resolution.}\label{fig:S_CgLp_exp_Q3}
\end{figure}

For the second quantization level of \SI{80}{nm}, a reference amplitude of $\SI{30}{\mu m}$ was again applied at lower frequencies. The theoretical limit in this case is -51.5 dB. Here the theoretical and practical limit nicely agree because the noise level is small relative to the 80 nm resolution. Fig. \ref{fig:S_CgLp_exp_Q3} shows the experimentally deduced sensitivity function. The quantization induced performance degradation was measured to be from \SI{5}{Hz} up to \SI{60}{Hz} indicating that larger quantization levels results in greater degradation in performance. It can be seen that both TR show improvement, where TR - $f_c$ shows the expected increased peak of sensitivity function.

\subsubsection{Noise attenuation}
\begin{figure}
	\centering
	\includegraphics[trim = {1.5cm 0 1.5cm 0}, width=1\linewidth]{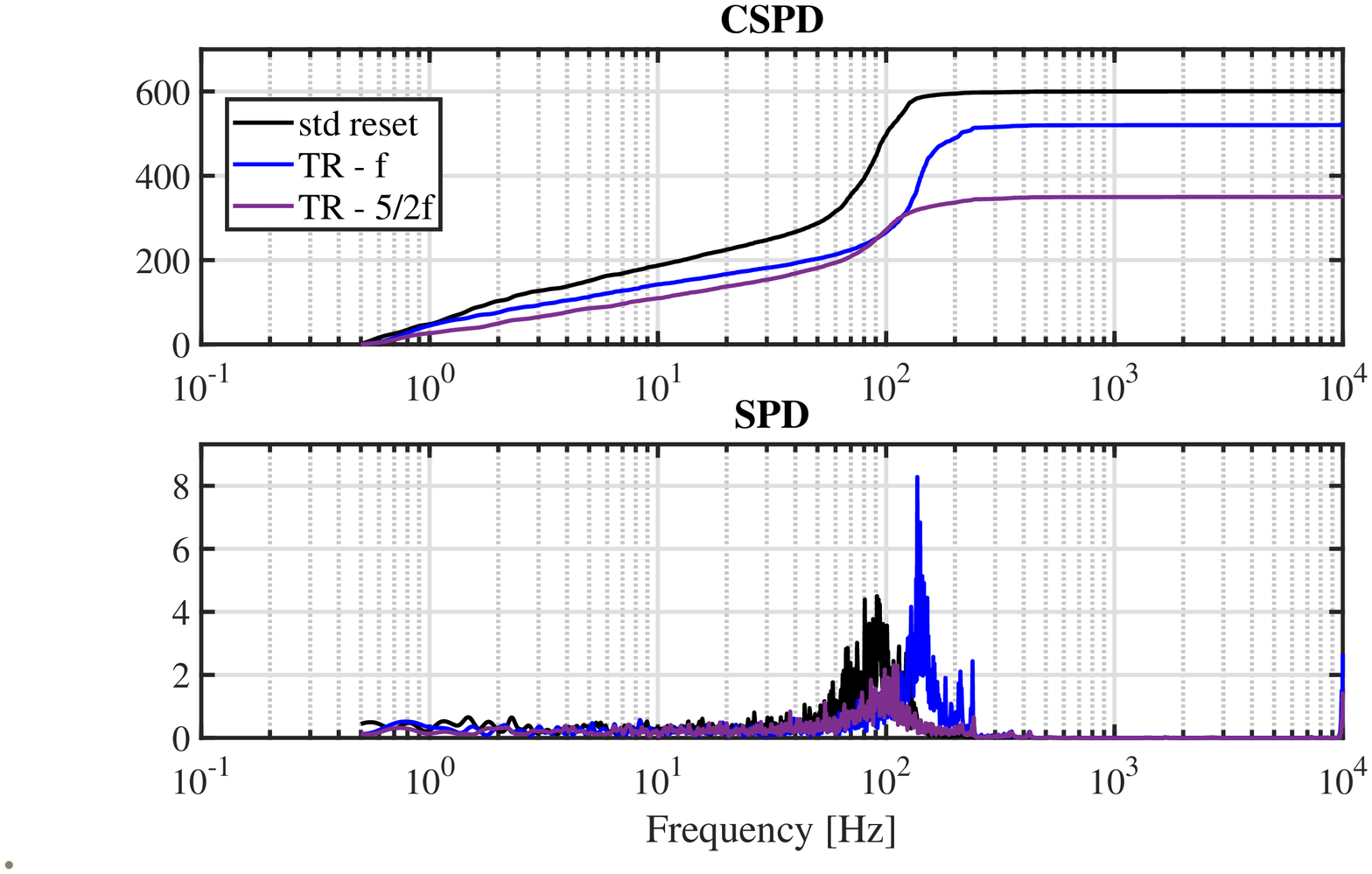}
	\caption{Experimentally deduced Cumulative Power Spectral Density function for CgLp-PID with and without Time Regularization (TR - $\frac{5}{2}\cdot f_c$ and TR - $f_c$). White noise with a maximum noise amplitude of \SI{700}{nm} is artificially added.}\label{fig:CSPD_combined_TR}
\end{figure}

Noise can increase the instances of excessive resetting. Time regularization influences the amount of resetting, therefore measurements have been performed to see the influence of time regularization on noise attenuation performance. Fig. \ref{fig:CSPD_combined_TR} shows the Cumulative Power Spectral Density (CPSD) of the error of standard reset and time regularization with white noise additionally added to the sensor signal. It can be seen that introducing a holding time can also improve the noise attenuation performance.

\section{Conclusion }\label{sec:Conclusion}
The most popular industrial controller, PID is limited by its linearity. Reset control is a promising alternative and can overcome Bode's gain-phase relationship. Reset control has been implemented in several practical applications, but no research has been done on the influence of quantization on its performance. In this paper it is shown that quantization can degrade the performance of reset control.
Time regularization was proposed with the novel purpose as a solution for quantization induced performance degradation. Tuning guidelines were provided, along with robustness analysis.
The influence of time regularization on noise attenuation was experimentally measured to be beneficial. Overall, time regularization is a promising solution within reset control to ensure that this nonlinear strategy is more widely applicable across different sensors with different levels of resolution in the motion and process control industry. Additional research to understand the effect of superposition of multiple reference signals on time regularized reset controllers can be considered in the future.

\bibliographystyle{IEEEtran}
\bibliography{library}

\end{document}